\documentclass[secnumarabic,amssymb, nobibnotes, aps, prd]{revtex4}%
\usepackage{graphicx}
\usepackage{dcolumn}
\usepackage{bm}
\usepackage{amsmath}%
\usepackage{amsfonts}%
\usepackage{amssymb}
\begin{document}

\title{Inter-galactic gravitational field effect on the propagation of light.}

\author{C. E. Navia, C. R. A. Augusto, D. F. Franceschini and K. H. Tsui}
\address{Instituto de F\'{\i}%
sica, Universidade Federal Fluminense, 24210-346,
Niter\'{o}i, RJ, Brazil} 

\date{\today}

\begin{abstract}
Recent observations of the luminosity-redshift in Type Ia supernovae suggest an accelerated inflation of the Universe as well as the observed matter density  showed to be less than the critical one, suggesting that a large fraction of the energy density of the Universe is in the form of dark energy with negative pressure (to supply repulsive forces). We present here an alternative mechanism on the basis of the photon energy loss in the inter-galactic gravitational field, and it is close to the Shapiro effect.  
It is argue that the redshift observed in distant galaxies is a cumulative process, a dominant redshift due to the  
Doppler effect plus a redshift due to photon energy loss in the intergalactic gravitational field.  We show that the last mechanism when interpreted as Doppler effect, supplies a non lineal relationship among the speed of recession of the galaxies and their distances. The effect is very tiny, even so, it increases with the distance and can be the key to explain the anomalous redshift observed in distant supernovae without the accelerated inflation hypothesis
\end{abstract}


\maketitle

The existence of galaxies (initially identified as nebulae) outside the Milk Way by Slipher \cite{slipher15} and Hubble
\cite{hubble31} in the first decades of the last century, opens the doors for the discovery of the proportionality constant between the recessional velocity of galaxies and their distances, called as Hubble's constant. However, Hubble himself did not agree that the redshift were only caused by the Doppler effect. In addition, there are some evidences that quasars and the so called companion galaxies present considerably higher redshift than Doppler effect \cite{fredman85, arp90}. Even so, the Hubble's law interpreted as Doppler effect is probably the most strong evidence in favor of the Big Bang Cosmology.

Recently a number of observations such as the distant supernovae of type Ia\cite{leibunggut01}, gives velocities  greater than the expected by the linear application of the Hubble constant, suggesting an accelerate inflation to the Universe (for a review see ref.\cite{olive06}). In addition, if the Universe is flat as WMAP's data suggests \cite{sanches06}, it is expected that $\Omega_m=\rho_m/\rho_c\sim 1$, where $\Omega_m$ is the ratio of the matter density to the critical density, where the critical density $\rho_c$ is defined as the density need to stop the expansion
\begin{equation}
\rho_c=\frac{3H_0^2}{8\pi G},
\end{equation}
where $H_0$ is the Hubble's constant and $G$ is the gravitational Newton's constant.
However, resent measurement of the cosmic background radiation has shown that
$\Omega_m$ is significantly less than unity, $\Omega_m\sim 0.231\pm0.021$ as obtained by WMAP \cite{sanches06}.

A key problem is to explain: a) why the density $\Omega_m<1$ nearly coincides with $\Omega_{total}=1$ as expected by inflation and b) which one is the mechanism that accelerates the inflation?
The most accepted scenario to solve the problem is to invoke a dominant ``dark energy'' component and whose equation of state is $w=p/\rho$ where $p$ (pressure) must be negative (to give repulsive forces). There are several candidates for the missing energy such as the
vacuum energy density, represented by the cosmological constant $\Lambda$ for which $w=-1$ and other cosmologies
based on a mixture of dark matter and quintessence\cite{zlatev99,picon00}, in all cases $(-1<w\leq 1))$. 

Evidently the above scenarios present some open questions such as, if there is dark energy it is just the a cosmological constant, and if so why is it so small?, why is the missing energy density so small compared to typical particle scales? If $\Omega_m \sim 0.2-0.3$ it is required a new mass scale 14 or so orders of magnitude smaller than the electroweak scale. 

On the other hands, we argue here that the  redshift observed in the radiation emitted by distant objects is a cumulative result of several process such as the gravitational redshift, Doppler effect and photon energy loss in the inter-galactic gravitational field, we will show here that the gravitational redshift supplies a very small redshift effect and can be negligible. While the dominant effect is the Doppler mechanism with a small contribution to the redshift due to the photon energy loss. The photon energy loss process comes from observation that
the speed of light depends on the strength of the gravitational potential along its path. This effect
is know as the Shapiro delay effect \cite{shapiro}, who first observed the time delay due to the solar gravitational field of radar signals returned from the surface of planets Venus and Mercury.
The contribution (to the delay) from the change in path is negligible.
The Shapiro effect constitute more a successful test of the general theory of gravitation. 

In this paper, the Shapiro effect is introduced in a different context. In the Shapiro's paper the subject is to obtain an expression for the difference between the proper time delay predicted in general relativity and the corresponding flat-space value. While here it is considered as a mechanism that can  contribute to the redshift from distant objects due to photon energy loss in the inter-galactic gravitational field as a cumulative effect on the dominant Doppler effect. We show that this mechanism when interpreted as a Doppler effect supplies a non-linear relationship among the speed of recession of the galaxies with their distances. The contribution of the photon energy loss effect is very tiny, even so, its increase with the distance and can be the key to explain the anomalous redshift observed in distant supernovae without the accelerate inflation hypothesis. 

%%%%%%%%%%%%%%%%%%%%%%%%%%%%%%%%
We begin our analysis with some remarks about local gravitation with spherical symmetry and a posterior extrapolation for a part of a structure on the very largest scale, that is a part of the Universe. This is possible, because recent observation on the Universe structure by WMAP has shown that the Universe in large scale is close to a flat geometry and almost smooth. 

The exact solution for the space-time metric surrounding a spherically symmetrical mass was obtained by 
Schwarzschild \cite{landau}
in 1916 and can be written as
\begin{equation}
ds^2=-\left(1-\frac{2GM}{c^2r}\right)dt^2+r^2(d\theta^2+\sin^2 \theta d\phi^2)+\left(1-\frac{2GM}{c^2r}\right)^{-1}dr^2,
\end{equation}
where the first relation on the right is known as the $g_{00}$ metric component
\begin{equation}
g_{00}=-\left(1-\frac{2GM}{c^2r}\right).
\end{equation} 
For light, we have $ds^2=0$ and the quantity ($1/-g_{00}$)  behaves as a refraction index \cite{fox76}
\begin{equation}
n=\frac{1}{-g_{00}}\approx \left(1+\frac{2GM}{c^2r}\right).
\end{equation}
The greatest value of $2GM/rc^2$ in the solar system is about $4\times 10^{-6}$ on the surface of the sun.
This means that the refractive index at solar limb is $n=1+4\times 10^{-6}$. This value of $n$ gives an angular deflection of the ray of light that just grazes the solar limb as  
\begin{equation}
\delta \phi \sim \frac{4GM}{rc^2}=1.75\;as.
\end{equation}
We show here that a consequence of considering this behavior of the refraction index is the energy loss of light during its propagation. The effect is very tiny, even so, it is measurable in cosmological scale. The result is a redshift due to photon energy loss and that when is added to the Doppler redshift, it gives a larger value than the one foreseen by the Hubble's law. This means that the ``observed'' accelerate inflation is only apparent. 

As also already has been commented, the WMAP data is close with a flat Universe
with an average density $\Omega_m=\rho_m/\rho_c=0.231\pm 0.021$ \cite{sanches06}. In addition, both the COBE and WMAP data \cite{scott06} have shown that the main deviation of the cosmic microwave background radiation (CMBR) from an isotropic distribution is due to Earth (solar barycenter) motion and produces a Doppler shift like dipole temperature, with an amplitude level of $\Delta T/T\sim 10^{-3}$. If this dipole is removed
the variations in the temperature of the CMBR is like a quadrupole with a level of $\Delta T/T <10^{-5}$. This result is considered as an evidence of that the Universe in large scale is approximately homogeneous and isotropic, an almost smooth Universe. In short, the Euclidean geometry can be used in the analysis.

Under this assumption, we can considered two point in the Universe. A Galaxy A at point A (as the light source) surrounded by other galaxies and a point B (as the detector) located at a distance $r$ from the galaxy A. 
In Fig.1, a part of the Universe is schematically represented.
If we chose the point A as origin of coordinates, for reasons of symmetry, the light rays will spread in the radial direction, following a straight line, consequently, $\theta = $constant and $\phi=$constant, or $d\theta = d\phi =0$. 
The galaxy A is at the center of a sphere of radii $r$, and only the mass M due to other objects
(galaxies) inside of this sphere contributes to the photon energy loss.

The speed of light at the emission (point A) is $c$. While the light arrives in the point B with a speed
\begin{equation}
v_{light}=\frac{c}{n}=\frac{c}{1+2GM/rc^2},
\end{equation} 
where $M$ is the mass inside the sphere of radius r and can be written as $M=\rho_m (4/3)\pi\;r^3$. Under this assumption, $v_{light}$ at point B can be written as 
\begin{equation}
v_{light}=\frac{c}{n}=\frac{c}{1+ \eta r^2},
\end{equation} 
where $\eta$ is given by
\begin{equation} 
\eta=\frac{8 \pi G\rho_m}{3c^2}=\frac{8\pi G \Omega_m \rho_c}{3c^2}. 
\end{equation}
Thus, the light emitted by the galaxy A with frequency $\nu_0$ has a lower frequency $\nu$ when it reaches the point B due to energy loss, and the ratio is
\begin{equation}
\frac{\nu}{\nu_0}=\frac{v_{light}^2}{c^2}=\left(\frac{1}{1+\eta r^2}\right)^2.
\end{equation}
On the other hand, the expression for the Doppler effect is \cite{fox76}
\begin{equation}
\frac{\nu}{\nu_0}=\frac{\sqrt{1-\beta}}{\sqrt{1+\beta}},
\end{equation}  
and represent a source moving away at a velocity $\beta(=v_s/c)$. Now, if the frequency shift due to photon energy loss is interpreted as Doppler effect, from Eq.9 and Eq.10 is possible to obtain the relation  
\begin{equation}
\frac{\nu}{\nu_0}=\frac{\sqrt{1-\beta}}{\sqrt{1+\beta}}=\left(\frac{1}{1+\eta r^2}\right)^2.
\end{equation}
Consequently, it is possible to obtain a fictitious correlation between
the distance and the apparent velocities of the galaxies as
\begin{equation}
v_s=c\left(\frac{(1+\eta r^2)^4-1}{(1+\eta r^2)^4+1}\right).
\end{equation}
This nonlinear correlation between velocity $v_s$ and distance $r$ can be the key to explain the apparent accelerated expansion of the universe. In Fig.2, the dashed line represents the Hubble's linear correlation with 
$H_0=72\pm 3\;km\;s^{-1}Mpc^{-1}$ \cite{pdg06}, and the dotted line represents the energy loss of photons in the inter-galactic gravitational field. The calculation is made for $\Omega_m=0.231\pm 0.021$ and the solid line represent the cumulative (addition) result between the Doppler effect and the photon energy loss effect respectively. In order to see the region where the type Ia supernovae are located in the Hubble's diagram, the velocity of recession $v_s$ of the objects can be expressed as a function of the redshift as
\begin{equation}
z= \frac{\sqrt{1+\beta}}{\sqrt{1-\beta}}-1,
\end{equation}
Putting Eq.12 upon Eq.13 is obtained the redshift $z$ as a function of the distance 
\begin{equation}
z= \eta r^2(2+\eta r^2).
\end{equation}
 The type Ia supernovae in the Hubble diagram reaches $z\approx 0.4-0.8$. In our scheme, only a fraction of this redshift correspond to the Doppler effect, there is also a contribution to the redshift due to photon energy loss process and that this being also interpreted as Doppler effect. The lower picture on Fig.2 summarized the situation.

On the other hand, in this framework is possible also to estimate the gravitational redshift effect as \cite{cuesta04}
\begin{equation}
z=\frac{1}{\sqrt{1-\frac{2GM}{rc^2}}}-1=\frac{1}{\sqrt{1-\eta r^2}}-1,
\end{equation}
this gravitational redshift increases as r increase. Even so, it is of the order of $z\sim 0.015$ even at cosmological distances larger than $r>1500\;Mpc$ from the source. In general the gravitational redshift is one order of magnitude smaller than the redshift due to photon energy loss process and it is negligible in the present analysis.

According to our results the accelerated inflation is only apparent, and it is not necessary to invoke the dark energy with a negative pressure in its state equation to take into account of the phenomenon. Evidently our result introduces several inconsistencies in the inflation, and important questions remain. For example, the observed low average density of the Universe with $\Omega_m\sim 0.2-0.3$, without dark energy, it appears to deny the possibility of a flat universe, as is suggested by WMAP data.  
Without dark energy it is expected an age of the Universe smaller than the actual estimation    
of $\sim 13.7$ Gyr. Is the new age of the Universe compatible with the necessary time for the formation of great structures in the universe? Does this smaller age of the Universe agree with the age estimate through the abundance of elements?  

On the other hand, the quintessence was introduced in the Greek cosmology by Aristotle where the Universe had four elements: earth, water, air and fire, plus an additional all-pervasive component, that accounted for the motion of star, Moon, and planets, called the quintessence of the universe, the ``ether''. The contemporary ether theory had its beginning in the Victorian era, with the Maxwell electromagnetic equations. Maxwell modelled the electro-magnetic waves as strands of ``ether'' or ``luminiferous ether'' through which light waves traveled. 

However, in the first decade of the twentieth century, chiefly as a result of the failure of attempts to observe the earth's motion relative to the ether, and the acceptance of the principle that such attempts must always fail, the ether was placed aside, especially after Einstein \cite{einstein52} wrote in his famous 1905 paper: $the\;luminiferous\;ether\;is\;superfluous$.
Nowadays there is no question that the ``vacuum'' has complex physical properties, notion like vacuum polarization is usual and this concept is the same than the original Maxwell's concept of the ether polarization. 
Currently, the ether or quintessence has been resurrected by the Big Bang cosmology in another context as the most promising candidate to explain the accelerated expansion of the Universe. Evidently our result is a challenge to this interpretation and the question remains. Is the dark energy superfluous?

\newpage

\begin{figure}[th]
\vspace*{2.0cm}
\hspace*{-1.0cm}
\includegraphics[clip,width=0.7
\textwidth,height=0.8\textheight,angle=0.] {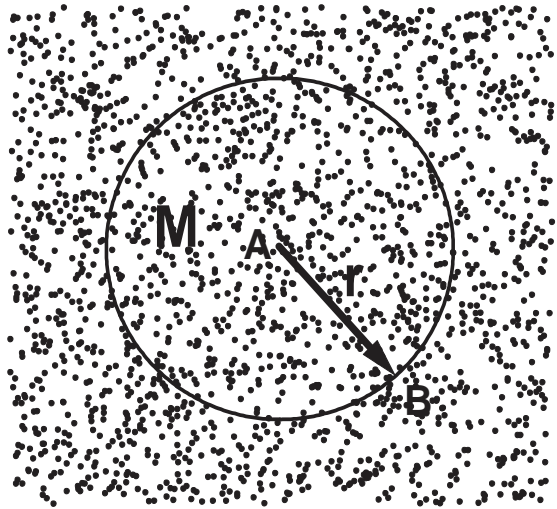}
\vspace*{-9.0cm}
\caption{Schematic representation of astronomical objects (galaxies) homogeneous and isotropically distributed in a part of the Universe (each point represent a galaxy). In the analysis is considered the galaxy A as light source and the point B at a distance $r$ from A, as detector, the mass $M=\rho_m (4/3)\pi r^3$ due the presence of all objects inside of the sphere of radii $r$ is obtained considering the Universe matter density $\rho_m=constant$.}%
\end{figure}

\begin{figure}[th]
\vspace*{2.0cm}
\hspace*{-0.0cm}
\includegraphics[clip,width=0.6
\textwidth,height=0.6\textheight,angle=0.] {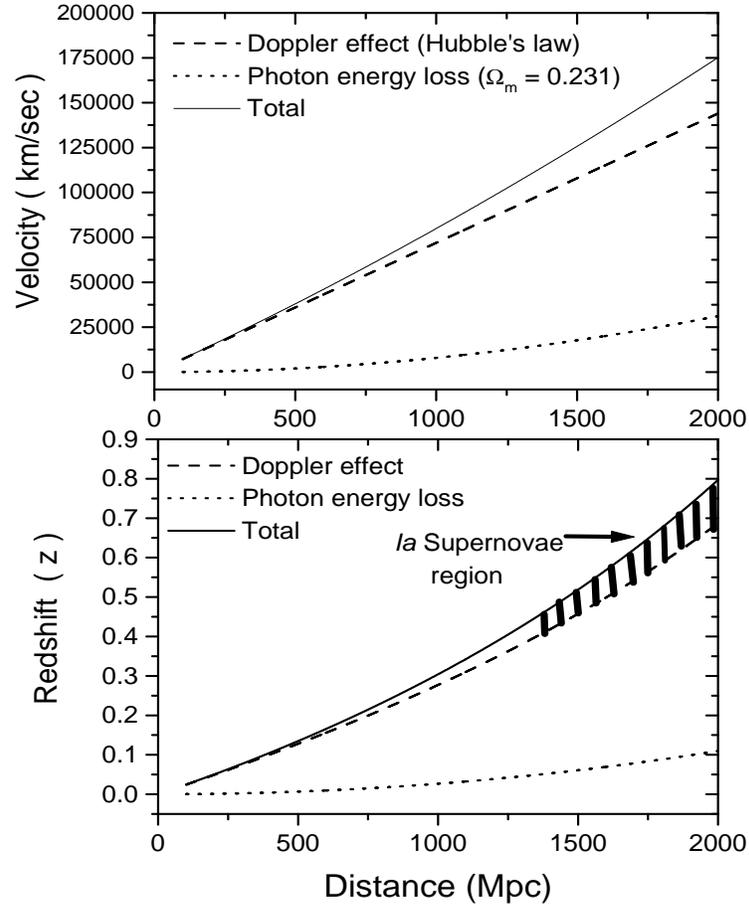}
\vspace*{-1.0cm}
\caption{Upper picture: Correlation among the velocity the recession of astronomical objects (galaxies) versus their distances. The dashed line represent the linear Hubble's law ($v_s=H_0\;r$), the dotted line represent the redshift due the photon energy loss process and interpreted as a fictitious Doppler effect and the solid line represent the addition of the two processes.
Lower picture: The same as upper picture, where the velocity of recession was written as function of the redshift parameter
($z$). The marked region represent the region where the type Ia supernovae are located. }%
\end{figure}

\end{document}